\documentclass[article,reprint,amsmath,amssymb,superscriptaddress,longbibliography]{revtex4-1}

\usepackage{graphicx}
\usepackage{dcolumn}
\usepackage{bm}
\usepackage{hyperref} 
\usepackage{breakurl}
\usepackage{multirow}
\usepackage{enumitem}
\usepackage{color}
\usepackage[pagewise]{lineno}
\setcitestyle{super} 
\usepackage{lineno}

\begin{document}


\title{Enhanced anisotropic superconductivity in the topological nodal-line semimetal In$_{x}$TaS$_{2}$ }

\author{Yupeng Li}
      \email{yupengLi@zju.edu.cn}
      \affiliation{Zhejiang Province Key Laboratory of Quantum Technology and Device, Department of Physics, Zhejiang University, Hangzhou 310027, China}
      \affiliation{State Key Laboratory of Silicon Materials, Zhejiang University, Hangzhou 310027, China}

\author{Zhongxiu Wu}
      \affiliation{Zhejiang Province Key Laboratory of Quantum Technology and Device, Department of Physics, Zhejiang University, Hangzhou 310027, China}

\author{Jingang Zhou}
      \affiliation{Zhejiang Province Key Laboratory of Quantum Technology and Device, Department of Physics, Zhejiang University, Hangzhou 310027, China}

\author{Kunliang Bu}
      \affiliation{Zhejiang Province Key Laboratory of Quantum Technology and Device, Department of Physics, Zhejiang University, Hangzhou 310027, China}

\author{Chenchao Xu}
      \affiliation{Zhejiang Province Key Laboratory of Quantum Technology and Device, Department of Physics, Zhejiang University, Hangzhou 310027, China}

\author{Lei Qiao}
      \affiliation{Zhejiang Province Key Laboratory of Quantum Technology and Device, Department of Physics, Zhejiang University, Hangzhou 310027, China}

\author{Miaocong Li}
      \affiliation{Zhejiang Province Key Laboratory of Quantum Technology and Device, Department of Physics, Zhejiang University, Hangzhou 310027, China}

\author{Hua Bai}
      \affiliation{Zhejiang Province Key Laboratory of Quantum Technology and Device, Department of Physics, Zhejiang University, Hangzhou 310027, China}

\author{Jiang Ma}
      \affiliation{Zhejiang Province Key Laboratory of Quantum Technology and Device, Department of Physics, Zhejiang University, Hangzhou 310027, China}

\author{Qian Tao}
      \affiliation{Zhejiang Province Key Laboratory of Quantum Technology and Device, Department of Physics, Zhejiang University, Hangzhou 310027, China}

\author{Chao Cao}
      \affiliation{Department of Physics, Hangzhou Normal University, Hangzhou 310036, China}

\author{Yi Yin}
      \affiliation{Zhejiang Province Key Laboratory of Quantum Technology and Device, Department of Physics, Zhejiang University, Hangzhou 310027, China}
      \affiliation{Collaborative Innovation Centre of Advanced Microstructures, Nanjing University, Nanjing 210093, China}

\author{Zhu-An Xu}
      \email{zhuan@zju.edu.cn}
      \affiliation{Zhejiang Province Key Laboratory of Quantum Technology and Device, Department of Physics, Zhejiang University, Hangzhou 310027, China}
      \affiliation{State Key Laboratory of Silicon Materials, Zhejiang University, Hangzhou 310027, China}
      \affiliation{Collaborative Innovation Centre of Advanced Microstructures, Nanjing University, Nanjing 210093, China}

\date{\today}

\begin{abstract}
Coexistence of topological bands and charge density wave (CDW) in topological materials has attracted immense attentions
because of their fantastic properties, such as axionic-CDW, three-dimensional quantum Hall effect, etc.
In this work, a nodal-line semimetal In$_{x}$TaS$_{2}$ characterized by CDW and superconductivity is successfully synthesized,
whose structure and topological bands (two separated Wely rings) are similar to In$_{0.58}$TaSe$_{2}$.
A $2 \times 2$ commensurate CDW is observed at low temperature in In$_{x}$TaS$_{2}$,
identified by transport properties and STM measurements. Moreover, superconductivity emerges below 0.69 K,
and the anisotropy ratio of upper critical field [$\Gamma = H_{c2}^{||ab}(0)/H_{c2}^{||c}(0)$] is significantly enhanced compared to 2H-TaS$_{2}$, which shares the same essential layer unit.
According to the Lawrence-Doniach model, the enhanced $\Gamma$ may be explained by the reduced effective mass in $k_{x} - k_{y}$ plane,
where Weyl rings locate.
Therefore, this type of layered topological systems may offer a platform to investigate highly anisotropic superconductivity
and to understand the extremely large upper critical field in the bulk or in the two-dimensional limit.
\end{abstract}

\maketitle


\section{Introduction}
Topological nodal-line semimetals (TNLSMs) \cite{TNLSM_BurkovAA_PRB2011} have been attracting tremendous attentions
due to the closed loop of band crossing formed in momentum space. Typical TNLSMs have been experimentally reported in the so-called 112 systems (In$_{x}$TaSe$_{2}$ \cite{InTaSe2_YpLi} and PbTaSe$_{2}$ \cite{PbTaSe2_BianG_NatC}), LiFeAs structure (ZrSiS \cite{ZrSiS_SchoopLM_NatC}), PtSn$_{4}$ \cite{PtSn4_WuY_NatP} and so on.
Unlike the zero-dimensional nodal points in Dirac semimetals \cite{SrIrO3_Carter_PRB2012,Na3Bi_LiuZK_science14,Cd3As2_NatC_NeupaneM}
and Weyl semimetals \cite{WeylSemi_WHM_PRX,TaAsDingH_ARPESWSM,TaAs_Hasan_Science,NbPFA_XDF_CPL15,NbP_PRB_WZ,NMR_YPL_FP17},
the one-dimensional (1D) nodal lines can be protected by certain symmetry \cite{TNLS_FangC_CPB2016}
no matter whether the spin-orbital coupling (SOC) is included.
Several intriguing properties have been predicted and experimentally observed in TNLSMs,
such as drumhead surface sates \cite{PbTaSe2_BianG_NatC,TlTaSe2_BianG_PRB16,Ca3P2_ChanYH_PRB16},
anomalous quantum oscillations \cite{TNLSMphase_LiC_PRL2018},
three-dimensional quantum Hall effect (3D QHE) \cite{NodalLineSM3DQHE_Molina_PRL18},
and topological superconductivity \cite{PbTaSe2_GuanSY_SciAdv16}.
As regards searching for bulk topological superconductors (TSCs),
one of the employed strategies is to induce superconductivity in topological materials \cite{Yplisummary_AQT19}
through the application of high pressure \cite{Bi2Te3SC_ZhangJL_PNAS11,WTe2_DFKang_NatC,NbAs2SC_Liyp},
intercalation between layers \cite{CuxBi2Se3_HorYS_PRL10,PbTaSe2_BianG_NatC}, or chemical doping \cite{CuxBi2Se3_HorYS_PRL10,PbTaSe2_BianG_NatC}, etc.

\begin{figure*}[!thb]
\begin{center}
\includegraphics[width=7in]{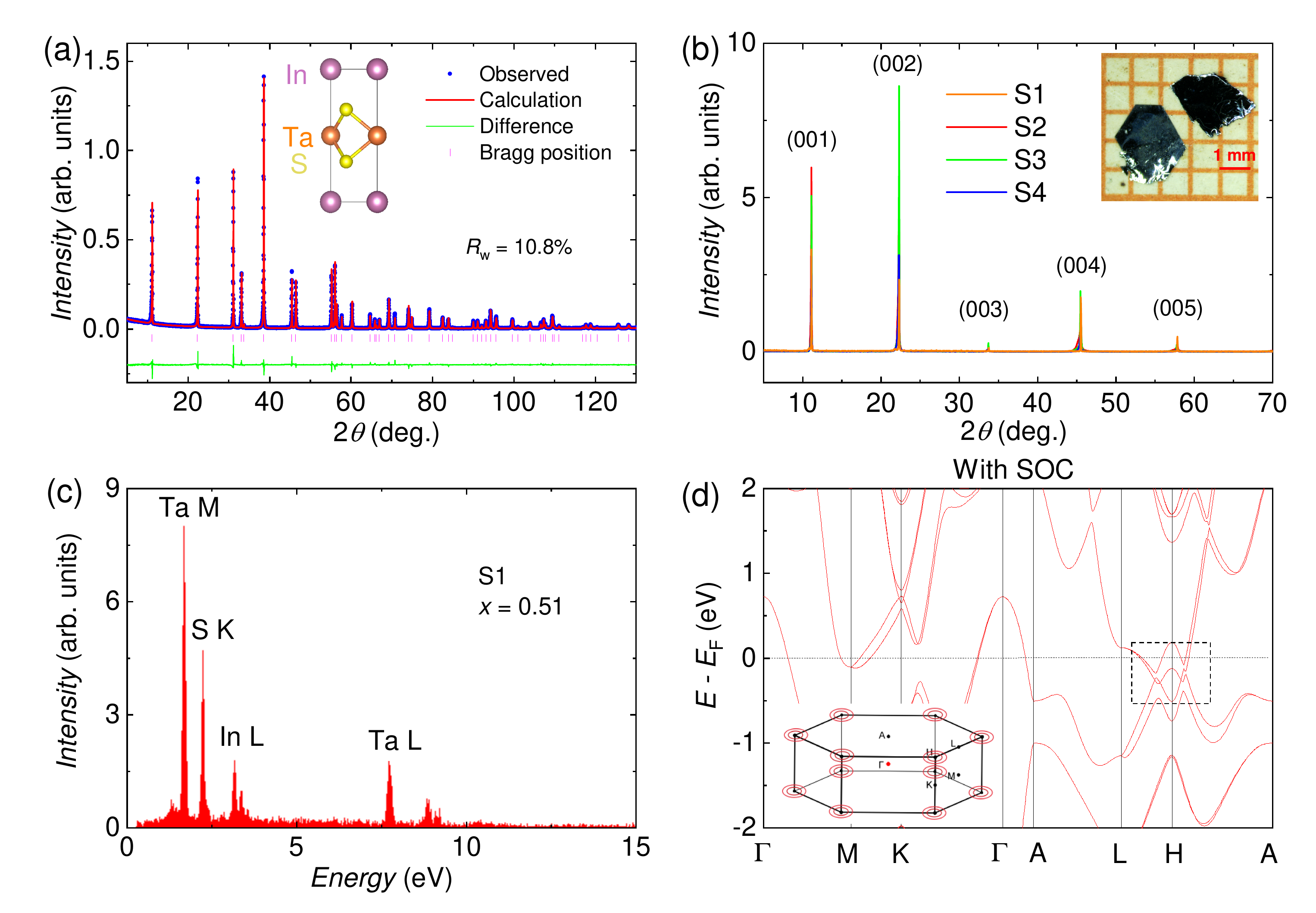}
\end{center}
\caption{\label{Fig1} (a) Rietveld refinement of powder XRD data for
polycrystalline InTaS$_{2}$. The inset is a side view of the InTaS$_{2}$ structure.
(b) XRD spectrum for the (00$l$) facet of single crystals In$_{x}$TaS$_{2}$.
The optical photograph of two selected samples are shown in the inset. (c) One of typical EDS spectrums collected on these
flat clean surfaces of single crystals. The In content $x$ is between 0.51 and 0.59. (d) Band structures of InTaS$_{2}$ with SOC.
Two separated Weyl rings appear at the H point in the first Brillouin zone, as shown in the inset.   }
\end{figure*}

Among these approaches, intercalation of atoms and molecules into the layered transition-metal dichalcogenides (TMDs)
$MX_{2}$ ($M$ is the transition metal, $X = S, Se, Te$) and other layered compounds is an effective one to significantly modify their properties.
For example, superconductivity can be induced in Cu$_{x}$TiSe$_{2}$ \cite{CuxTiSe2_MorosanE_NP06}
and Cu$_{x}$Bi$_{2}$Se$_{3}$ \cite{CuxBi2Se3_HorYS_PRL10}; the intercalated graphite exhibits more excellent electric
and optical features \cite{IntercalatedGraphite_Dresselhaus_AP02}, which greatly contributes to extensive applications;
the TNLSM PbTaSe$_{2}$ viewed as Pb atoms intercalation in TaSe$_{2}$
introduces not only topological bands, such as InTaSe$_{2}$ \cite{InTaSe2_YpLi}, TlTaSe$_{2}$ \cite{TlTaSe2_BianG_PRB16}, InNbS$_{2}$, and InNbSe$_{2}$ \cite{InTaSe2_DuYP_PRB17},
but also possible Majorana bound states in the superconducting vortices \cite{PbTaSe2_GuanSY_SciAdv16,PbTaSe2Fe_ZhangSTS_PRB2020}.
Moreover, the intercalated layered compounds possibly host the higher superconducting transition
($T_{c}$) or highly anisotropic superconductivity \cite{PbTaSe2SCP_ZhnagCL_PRB16,NaxTaS2_FangL_PRB05}.


In this paper, the In-intercalated TNLSM In$_{x}$TaS$_{2}$ hosting both superconductivity
and charge density wave (CDW) is successfully synthesized.
It has the same structure and similar topological bands as InTaSe$_{2}$ \cite{InTaSe2_YpLi},
whose two separate Weyl rings exist at the H point.
It is a little different that the only one $2 \times 2$ commensurate CDW (CCDW) remains.
The superconductivity is observed, and
the extremely large anisotropy ratio of upper critical field is obtained in four samples,
which may be related to the small effective mass in the $ab$ plane.

\begin{figure*}[!thb]
\begin{center}
\includegraphics[width=7in]{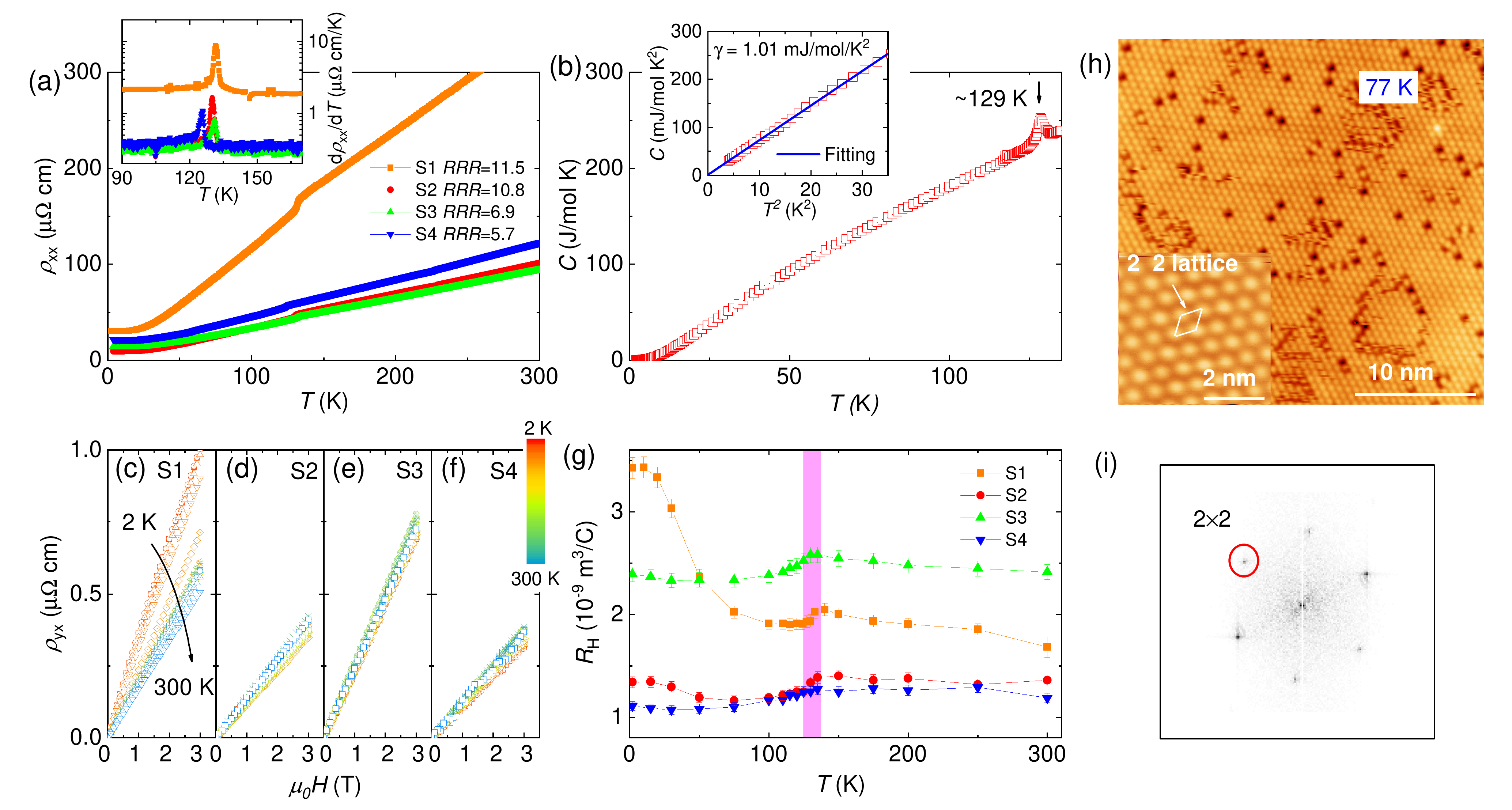}
\end{center}
\caption{\label{Fig2} (a) Electrical resistivity $\rho_{xx}$ of four In$_{x}$TaS$_{2}$ samples exhibiting a CDW-like transition.
The differential resistivity in the inset displays the transition temperatures $\sim$ 130 K.
(b) Low-temperature specific heat of In$_{x}$TaS$_{2}$ showing a distinct jump at $\sim$ 129 K and zero magnetic field.
The small $\gamma$ = 1.01 mJ mol$^{-1}$K$^{-1}$ is obtained in the inset. (c-f) Magnetic-field dependent $\rho_{yx}$ at different temperatures.
(g) Hall coefficients of four samples. The magenta range marks the transition.
(h) STM images (V$_{b}$ = -1V, I$_{t}$ = 20 pA) of the sample surface at 77 K.
An enlarged range of a perfect surface is shown in the inset (V$_{b}$ = 1V, I$_{t}$ = 100 pA),
implying the 2$\times$2 superlattice. (i) FFT image of (h). The $2 \times 2$ CCDW is marked by the red circle.  }
\end{figure*}

\begin{figure*}[!thb]
\begin{center}
\includegraphics[width=7in]{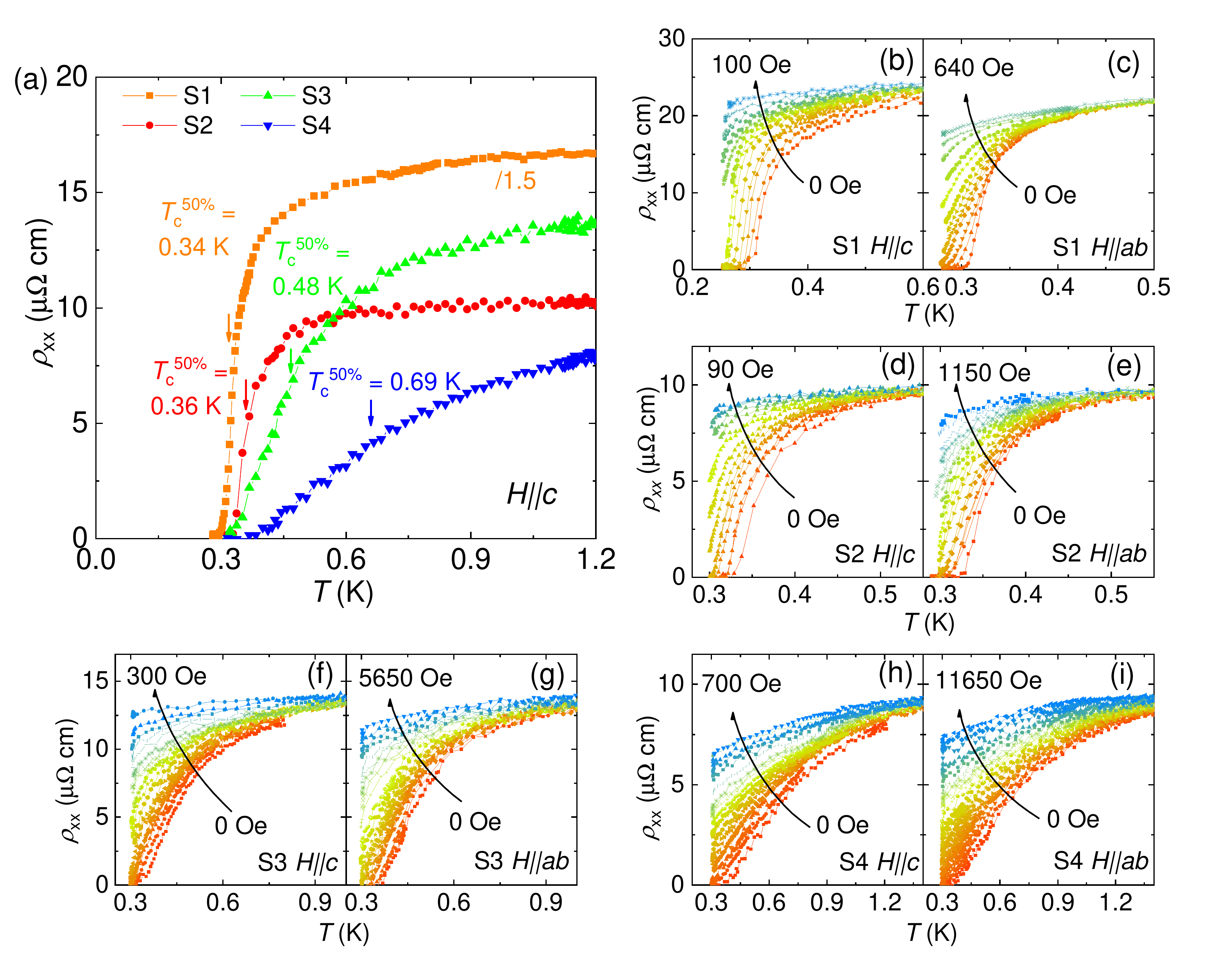}
\end{center}
\caption{\label{Fig3} (a) $\rho_{xx}$ of samples S1, S2, S3, and S4 showing the superconducting transition
at $T_{c}^{50\%}$ = 0.34 K, 0.36 K, 0.48 K, and 0.69 K, respectively. $T_{c}^{50\%}$ is determined by 50\% drop of the normal-state resistivity.
(b-i) Temperature dependence of in-plane and out-of-plane resistivity for four samples.   }
\label{Fig4}
\end{figure*}

\section{Experiment}

The polycrystalline InTaS$_{2}$ and single crystals In$_{x}$TaS$_{2}$ were prepared using the
solid-state reaction method and the vapor transport method, respectively, also referred to In$_{x}$TaSe$_{2}$\cite{InTaSe2_YpLi}.
The x-ray diffraction (XRD) data were collected using a monochromatic Cu K$_{\alpha1}$ radiation. An energy-dispersive x-ray spectroscopy (EDS)
was employed to analyze chemical compositions of samples. A standard six-probe technique was carried out
to measure the longitudinal resistivity and Hall resistivity on an Oxford $^{3}$He-based cryostat
and a physical properties measurement system (PPMS). Scanning tunneling microscopy (STM) measurements were performed in
a commercial unisoku-UHV1500S STM system. The samples were cleaved in situ
at $\sim$77 K, then inserted into the STM measurement stage.

The density function theory (DFT) calculations were performed using the generalized gradient
approximation (GGA) method and the Perdew-Burke-Ernzerhoff (PBE) exchange correlation functional. The lattice constants and
the atomic coordinates were used from Rietveld-refined XRD data. A 18$\times$18$\times$6
Monkhorst-Pack $k$-point mesh and SOC were applied in the computations.

\section{Results}

\subsection{Structure and topological bands}

InTaS$_{2}$ has the same noncentrosymmetric structure \emph{P$\bar{6}$m}2
as InTaSe$_{2}$ \cite{InTaSe2_YpLi},
as shown in the inset of Fig. 1(a), the side view of the crystal structure.
This structure can be well verified by the Rietveld refinement of polycrystalline
powder XRD data [Fig. 1(a)]. The both reliable factor $R_{wp}=$ 10.8\% and small difference
between observed data and calculations illustrate the good refinement.
The refined lattice constants
are $a$ = $b$ = 3.3290 \AA \ and $c$ = 7.9891 \AA.
High-quality single crystals In$_{x}$TaS$_{2}$ with various In content
can be grown by the vapor transport method,
and the plate-like samples are obtained as large as 3 mm$\times$2 mm [inset of Fig. 1(b)].
Their XRD spectrums for the (00$l$) facet are collected in Fig. 1(b), suggesting the good single crystal quality.
The grown single crystals usually have a large amount of In vacancy.
A typical EDS pattern of a In$_{x}$TaS$_{2}$ single-crystalline sample S1 is shown in Fig. 1(c),
in which the chemical composition is In:Ta:S = 0.51:1:2.
The $x$ value of four In$_{x}$TaSe$_{2}$ samples varies between 0.51 and 0.59,
and phase separation easily emerges beyond this $x$ range.

Fig. 1(d) shows the band structure of InTaS$_{2}$ with the inclusion of SOC obtained by the DFT calculations.
The main features are quite similar to the other 112 systems, such as InTaSe$_{2}$ \cite{InTaSe2_YpLi}, InNbS$_{2}$ \cite{InTaSe2_DuYP_PRB17},
PbTaSe$_{2}$ \cite{PbTaSe2_BianG_NatC}, and TlTaSe$_{2}$ \cite{TlTaSe2_BianG_PRB16}.
The band inversion exists at the H point due to the hybridization of a hole pocket
from Ta-5d orbitals and an electron pocket derived from In-5p orbitals.
When the mirror reflection with respect to the In atomic plane is taken into consideration,
these inversed bands are topological invariant. The four-fold-degenerate Dirac-type nodal ring
splits into a pair of two-fold-degenerate nodal rings (Weyl rings) at the H point in the presence of SOC,
as seen in the inset of Fig. 1(d). These Weyl rings remain gapless as a result of the symmetry protection,
and they locate at $E - E_{F} \sim -0.25$ eV, slightly below the Fermi level.
Interestingly, the In vacancy in In$_{x}$TaS$_{2}$ is supposed to shift the Fermi level
down to the Weyl rings, which is also observed in In$_{0.58}$TaSe$_{2}$ \cite{InTaSe2_YpLi}.
Each Weyl ring possesses a Berry phase of $\pi$, and they can be
connected by drumhead surface states, a kind of nearly flat bands,
which may exhibit a van Hove singularity, as discussed in Refs[3,16,28,32]
\cite{InTaSe2_DuYP_PRB17,TlTaSe2_BianG_PRB16,PbTaSe2_BianG_NatC,Ca3P2_ChanYH_PRB16}.

\subsection{CDW states}

More fantastic features can be observed in the temperature dependence of
resistivity for In$_{x}$TaS$_{2}$, shown in Fig. 2.
Fig. 2(a) shows the residual resistivity ratios ($RRR$)
of samples S1, S2, S3, and S4 are 11.5, 10.8, 6.9, and 5.7, respectively.
All resistivity curves exhibit one sudden drop
at $\sim$ 130 K, a little different from two CDW transitions
in In$_{0.58}$TaSe$_{2}$ \cite{InTaSe2_YpLi}.
The transition temperatures can be identified by the differential resistivity in the inset of Fig. 2(a),
and also confirmed by the jump of specific heat at $\sim$ 129 K in Fig. 2(b).
The small $\gamma = 1.01$ mJ/mol/K$^{2}$ is obtained by fitting the low-temperature specific heat [inset of Fig. 2(b)].
In Fig. 2(c-f), the positive and linear Hall resistivity as a function of magnetic field ($H$) indicates
the dominated carrier is hole in this system. In Fig. 2(g),
the associated transitions of Hall coefficients ($R_{H}$) are also observed in the magenta range of temperature,
in agreement with the longitudinal resistivity and specific heat measurements.
Upon decreasing temperature, $R_{H}$ decreases at this transition point
in In$_{x}$TaS$_{2}$, different from the increase of $R_{H}$ in In$_{0.58}$TaSe$_{2}$ \cite{InTaSe2_YpLi}.
This behavior implies the possible distinct transition behavior or multiband feature, and the latter one is supposed to dominate here.
The Hall coefficient in the multiband system, such as In$_{x}$TaS$_{2}$, can be approximatively described by the two-band model
$R_{H}=[R_{H}^{h}(\sigma_{xx}^{h})^{2}-R_{H}^{e}(\sigma_{xx}^{e})^{2}]/(\sigma_{xx}^{h}+\sigma_{xx}^{e})^{2}$
\cite{W2As3_LiYP_PRB18}, where $R_{H}^{h}$ and $R_{H}^{e}$ denote the Hall
coefficient for hole and electron, respectively, $\sigma_{xx}^{h}$ and $\sigma_{xx}^{e}$
are the hole conductivity and electron conductivity, respectively.
Therefore, the Hall coefficient of the multiband system changes much complicatedly,
especially in CDW systems, where band gaps emerge.

To further investigate this transition, we perform STM measurements
at liquid nitrogen temperature (77 K). In Fig. 2(h), we show a STM topography obtained
with a bias voltage V$_{b}$ = -1 V and a tunneling current $I_{t}$ = 20 pA, from which a triangular lattice can be observed.
The triangular lattice can be further discerned in an enlarged small-area topography [inset of Fig. 2(f)].
The distance between adjacent bright spots is 6.951 \AA \ ($\sim$ 2$a$,
$a$ is the lattice constant). The basic element of lattice is the 2$\times$2 superlattice,
instead of the 1$\times$1 atomic lattice. The superlattice is also confirmed in the fast Fourier-transformed (FFT) result in Fig. 2(i).
The pattern marked by the red circle represents the 2$\times$2 lattice,
with a wave vector $\frac{1}{2}\vec{Q}_{0}$ ($\vec{Q}_{0} = 4\pi/\sqrt{3}a$).
The absence of atomic lattice in STM topography is similar to that for 1T-TaS$_{2}$ \cite{1TTaS2_ChoD_NatC16,1TTaS2_BuKL_CommP19}.
The low-temperature state below the transition is thus a 2$\times$2 CCDW state.
This type of 2$\times$2 CCDW can also be observed in other intercalated
$MX_{2}$ \cite{FeTaS2_DaiZ_PRB93}, including In$_{0.58}$TaSe$_{2}$ \cite{InTaSe2_YpLi}.
In addition, the random black spots in Fig. 2(h) may stem from
In atoms, which are randomly exfoliated from the In layer when the sample is cleaved.

\begin{figure}[!thb]
\begin{center}
\includegraphics[width=3.5in]{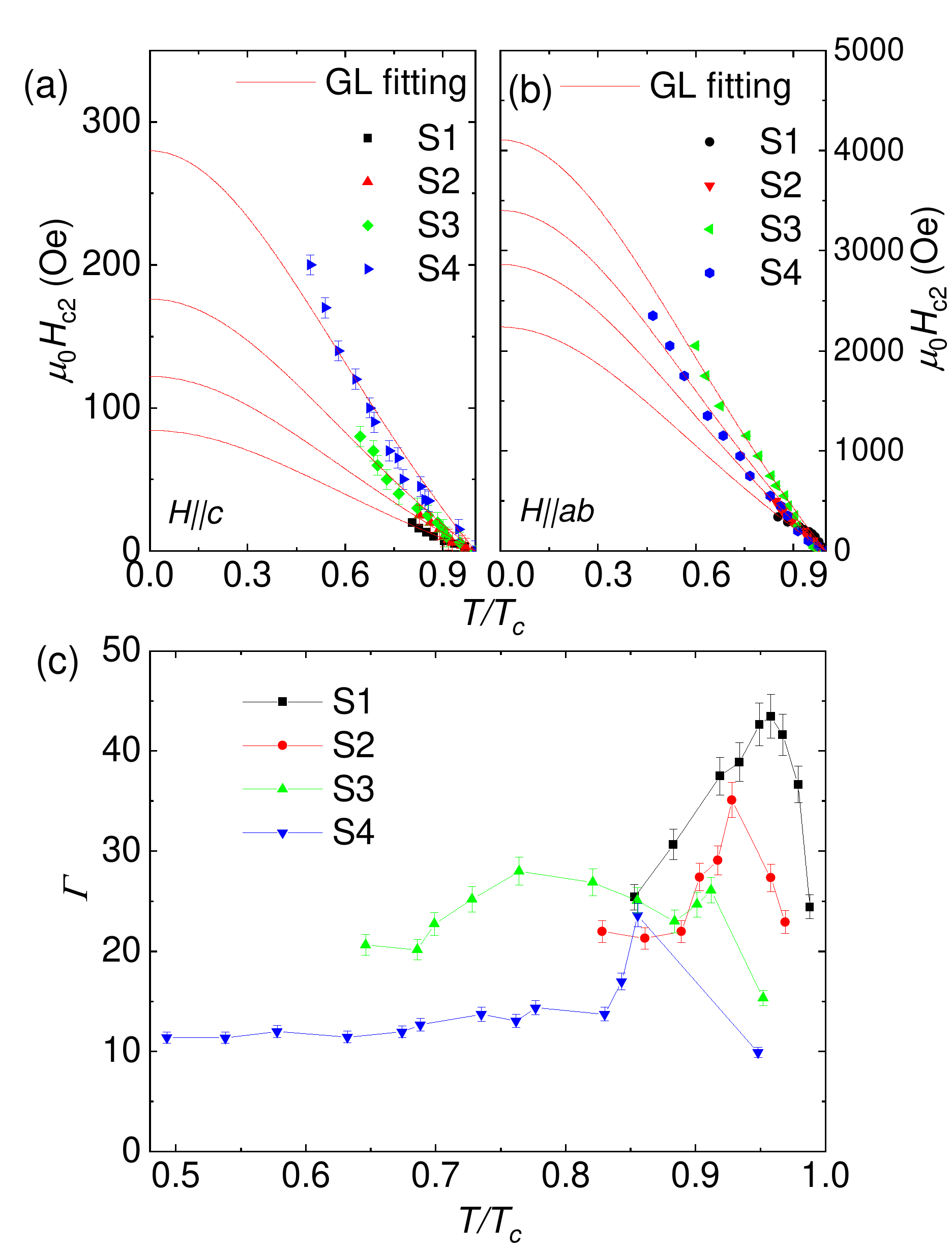}
\end{center}
\caption{\label{Fig4} Out-of-plane (a) and in-plane (b) upper critical fields with the GL fitting (solid line).
(c) The anisotropy ratio $\Gamma = H_{c2}^{||ab}/H_{c2}^{||c}$ for four samples,
all larger than 12. }
\label{Fig4}
\end{figure}

\subsection{Anisotropic superconductivity}

In Fig. 3, superconductivity of samples is observed at low temperature.
The superconducting transition temperature ($T_{c}^{50\%}$) is 0.34 K, 0.36 K, 0.48 K, and 0.69 K
for samples S1, S2, S3, and S4, respectively, which is extracted by 50\% drop of normal-state resistivity.
$T_{c}$ in In$_{x}$TaS$_{2}$ is a little smaller than $T_{c} \sim 0.8$ K in
2H-TaS$_{2}$ with the 3$\times$3 CCDW at 78 K \cite{2HTaS2_MorrisRC_PRB73},
and also smaller than $T_{c} = 0.91$ K for In$_{0.58}$TaSe$_{2}$ with
two CDW transitions (2 $\times \sqrt{3}$ CCDW at 117 K and $2 \times 2$ CCDW at 77 K) \cite{InTaSe2_YpLi}.
The lower $T_{c}$ for In$_{x}$TaS$_{2}$ may be due to its higher $2 \times 2$ CCDW transition
(130 K), which is higher than In$_{0.58}$TaSe$_{2}$. This is consistent with the 
typical phase diagram of CDW superconductors, i.e., Cu$_{x}$TiSe$_{2}$ \cite{CuxTiSe2_MorosanE_NP06},
in which $T_{c}$ is enhanced upon suppressing CDW.
The widths of superconducting transitions for samples S1 and S2 are much narrower
than samples S3 and S4, indicating higher sample quality.
The temperature dependence of resistivity under different magnetic fields are shown in Fig. 3 (b-i)
for both $H$ applied in the $ab$ plane and along the $c$-axis.

\begin{table}[!thb]
\tabcolsep 0pt \caption{\label{SC} Summary of physical parameters for the four In$_{x}$TaS$_{2}$ samples.     } \vspace*{-12pt}
\begin{center}
\def\temptablewidth{1.0\columnwidth}
{\rule{\temptablewidth}{1pt}}
\begin{tabular*}{\temptablewidth}{@{\extracolsep{\fill}}ccccc}
Sample                   &S1          &S2       &S3     &S4  \\ \hline
$x$                      &0.51        &0.59     &0.56   &0.58\\
$c$ (\AA)                &7.9647      &7.9753   &7.9773 &7.9849\\
$RRR$                    &11.5        &10.8     &6.9    &5.7\\
$T_{CDW}$ (K)            &132         &130      &131    &126\\
$T_{c}^{50\%}$ (K)       &0.34        &0.36     &0.48   &0.69\\
$H_{c2}^{||c}$(0K) (Oe)  &84          &122      &176    &280\\
$H_{c2}^{||ab}$(0K) (Oe) &2238        &2862     &4106   &3403\\
$\Gamma$ (0K)            &26          &23       &23     &12 \\
$m_{c}/m_{ab}$           &676         &529      &529    &144 \\
$\xi_{c}$ (nm)           &7.4         &7.0      &5.9    &8.9\\
$\xi_{ab}$ (nm)          &197.9       &164.3    &136.7  &108.5 \\
$n$ (2K) ($\times10^{27}$ m$^{-3}$) &1.82  &4.66 &2.61 &5.63    \\
$\rho_{0}$ ($\mu\Omega cm$) &25.1     &9.3      &13.6  &9.1\\
$k_{F}$ ($\times10^{9} m^{-1}$) &3.3       &4.5  &3.7  &4.8 \\
$l$ (nm)                  &29.8       &42.7      &43.2 &38.7\\
\end{tabular*}
{\rule{\temptablewidth}{1pt}}
\vspace*{-18pt}
\end{center}
\end{table}

\vspace{3ex}

The upper critical fields $H_{c2}$ as a function of $T_{c}^{50\%}$
are summarized in Fig. 4(a) and (b), and approximatively fitted by the
Ginzberg-Landau (GL) model (solid lines), $H_{c2}(t) = H_{c2}(0)(1 - t^{2})/(1 + t^{2})$,
where $t = T/T_{c}$. Taking the sample S1 for example, $H_{c2}^{||ab}$(0) = 2238 Oe
and $H_{c2}^{||c}$(0) = 84 Oe, while the parameters of other samples are listed in Table \ref{SC}.
Further effective mass can also be obtained according to the Lawrence-Doniach
model \cite{Lawrence-DoniachModel1970,NdFeAsOF_JiaY_APL08}, and the anisotropy ratio
$\Gamma$ is given by the following relation
\begin{equation}\label{LD}
\Gamma = H_{c2}^{||ab}/H_{c2}^{||c} = (m_{c}/m_{ab})^{1/2} = \xi_{ab}/\xi_{c},
\end{equation}
where $m_{c}$ and $m_{ab}$ are the effective mass tensor along the $c$ axis and $ab$ plane,
respectively, and $\xi_{ab}$ and $\xi_{c}$ are the coherence length
along the $ab$ plane and $c$ axis, respectively.
The anisotropy ratio $\Gamma$ vs. $t$ of these samples are displayed in Fig. 4(c).
The resultant $\Gamma$ for samples are all larger than 12,
implying the highly anisotropic superconductivity.
The $\Gamma$ value estimated at the extrapolated zero temperature is $\sim$ 26, 23, 23, 
and 12 for samples S1, S2, S3, and S4 with different $RRR$, respectively,
much larger than the most layered compounds, such as 2H-TaS$_{2}$
\cite{2HTaS2_MorrisRC_PRB73} ($\Gamma = 6.7$ in Table \ref{ASC}),
typical iron-based superconductors (IBSs) [``122"-type (Ba,K)Fe$_{2}$As$_{2}$ ($\Gamma < 2 $) \cite{BaKFe2As2_YHQ_Nature09},
``11"-type Fe$_{1+y}$Te$_{0.6}$Se$_{0.4}$ ($\Gamma < 1.8$), ``1111"-type NdFeAsO$_{0.82}$F$_{0.18}$ ($\Gamma < 5$), ``1144"-type RbEuFe$_{4}$As$_{4}$
($\Gamma < 1.7$) \cite{RbEuFe4As4_SmylieMP_PRB18}, ``112"-type Ca$_{1-x}$La$_{x}$FeAs$_{2}$ ($\Gamma < 4.2$)
\cite{CaLaFeAs2_ZW_APE14}], etc. Several TMDs-related compounds are listed in Table \ref{ASC}.
Unfortunately, the guided relation between $T_{c}$ and layer distance $d$ or $\Gamma$ in these TMDs-related compounds seem not to be widely concluded.
However, intercalated layered compounds seems a good method to possess highly
anisotropic superconductivity and large $H_{c2}^{||ab}$ in bulk
state, even exceeding the Pauli limit $H_{p}$ \cite{NaxTaS2_FangL_PRB05,TaSe2_BH_JPCM18,MisfitNbSe_BaiH_JPCM18,misfitNbSe22_BaiH_MRE20}.

According to the anisotropic Ginzburg-Landau formulas,
$H_{c2}^{||c} = \Phi_{0}/2\pi\xi_{ab}^{2}$ and $H_{c2}^{||ab} = \Phi_{0}/2\pi\xi_{ab}\xi_{c}$,
where $\Phi_{0}$ is the flux quantum, the GL coherence lengths $\xi_{ab}$ and $\xi_{c}$
at zero temperature are calculated for four samples.
The coherence length $\xi_{c}$ perpendicular to the TaS$_{2}$ layer is
more than 7 times larger than the distance $d = c = 7.9647 \AA$ between TaS$_{2}$ layers (Table \ref{SC}),
illustrating that the superconductivity of In$_{x}$TaS$_{2}$ remains
three dimensional in nature.
The carrier density and $\rho_{0}$ at 2 K are estimated from $R_{H}$ and low-temperature resistivity, respectively.
The Fermi vector $k_{F}$
and the mean free path $\ell$ are approximately inferred from the relation
$\ell = \hbar k_{F}/\rho_{0}ne^{2}$ and $k_{F} = (2\pi^{2}n)^{1/3}$, respectively.
All the physical parameters for four samples are summarized in Table \ref{SC}.

\begin{table}
\tabcolsep 0pt \caption{\label{ASC} Comparison of physical properties and anisotropy ratio of several TMD-related superconductors.
Some compounds have two CDW transitions, which are denoted as $T_{CDW1}$ and $T_{CDW2}$, respectively.     } \vspace*{-12pt}
\begin{center}
\def\temptablewidth{1.0\columnwidth}
{\rule{\temptablewidth}{1pt}}
\begin{tabular*}{\temptablewidth}{@{\extracolsep{\fill}}ccccc}
Material                    &$T_{c}$ (K)  &$T_{CDW1}$ (K)    &$c$ (\AA)  &$\Gamma$   \\
                            &            &   ($T_{CDW2}$)                                                          \\ \hline
2H-TaSe$_{2}$ \cite{TaSe2_Yokota_2000}             &0.14    &90 (121)        &12.71           &2.6                 \\
PbTaSe$_{2}$ \cite{PbTaSe2SCP_ZhnagCL_PRB16}       &3.8     &--             &9.35     &$\sim$4               \\
In$_{0.58}$TaSe$_{2}$ \cite{InTaSe2_YpLi}          &0.91     &77 (117)        &8.3231         &4.6                \\
In$_{0.51}$TaS$_{2}$               &0.34    &132           &7.9647       &$>$12              \\
2H-TaS$_{2}$ \cite{2HTaS2_MorrisRC_PRB73}          &0.8     &78            &12.097      &6.7                \\
Na$_{0.1}$TaS$_{2}$ \cite{NaxTaS2_FangL_PRB05}     &4.3     &--              &12.082    &6.4                \\
Cu$_{x}$TaS$_{2}$ \cite{Cu0.03TaS2_ZhuXD_JPCM09}   &4.03    &55              &12.11        &5.1                 \\
\end{tabular*}
{\rule{\temptablewidth}{1pt}}
\vspace*{-18pt}
\end{center}
\end{table}

Due to the limit on the lowest temperature which we can reach in our measurements,
the intrinsic anisotropy ratio may be a little overestimated, but its value is still supposed to be very large ($> 10$).
Subsequently, a remarkably large effective mass ratio $m_{c}/m_{ab}$ ($\gg 100$) can be extracted from Eq. (\ref{LD}).
Taking into consideration of small $\gamma = m^{*}k_{F}k_{B}^2/3\hbar^{2} = 43.76$ J/m$^{3}$/K$^{2}$ from the Landau Fermi-liquid theory \cite{baym2008landau},
which is obtained from the specific heat at constant volume,
the geometric mean of effective mass $m^{*}$ is approximately 2.2 $m_{e}$, suggesting the possible small effective mass $m_{ab}$ in the $ab$ plane.
The decreased effective mass may result from the linear band crossings (Weyl rings),
which locate in the $ab$ plane near the Fermi level due to the In vacancy.

\section{Conclusions}

We systematically investigate anisotropic upper
critical field in a nodal-line semimetal In$_{x}$TaS$_{2}$.
Similar to In$_{0.58}$TaSe$_{2}$, CDW, nodal-line topological states, and
superconductivity coexist in In$_{x}$TaS$_{2}$. A $2 \times 2$ CCDW transition is
observed at approximately 130 K supported by STM and transport measurements,
and then superconducting transitions of four samples emerge in the temperature range between 0.34 K and 0.69 K.

Among these physical phases in this system, one of the interesting points is the
gigantic anisotropy of upper critical field, which is significantly larger than that of 2H-TaS$_{2}$.
Several origins may account for this property in 3D materials.
In IBSs, the anisotropy ratio $\Gamma$ appears to be related to the inter-layer coupling strength and the distance $d$
between the charge reservoir layers and the conducting layers \cite{CaLaFeAs2_ZW_APE14}.
In the FeSe system, the larger $\Gamma$ may result from the larger $d$, which is likely correlated with the higher $T_{c}$ \cite{LiFeSe_SunSS_PRB17}.
From the Lawrence-Doniach model, the anisotropic effective mass has an influence on the $\Gamma$ value.
Considering the lower $T_{c}$ in this In-intercalated TaS$_{2}$ system, the origin of
large $\Gamma$ may be different from the IBSs system.
We propose that the large $\Gamma$ in In$_{x}$TaS$_{2}$ may result from gigantic anisotropic effective mass,
because linear band crossings (Weyl rings) locate in the $ab$ plane and reduce the effective mass tensor in this plane.
The vacancy of In shifts Weyl rings much close to the Fermi level, and the superconducting gap may emerge in the Weyl rings as well.
Therefore, we suppose the In-induced Weyl rings in this 112 system may contribute to the enhanced anisotropic effective mass,
then generating the large superconducting anisotropy. This scenario may also be applied to the topological materials with similar band structures, such as In$_{x}$TaSe$_{2}$, PbTaSe$_{2}$, etc.

In addition, the correlation between the large anisotropic superconductivity and band crossing in topological materials deserves further investigation.
Whether the superconducting gap emerges in the Weyl rings and the possible topological superconductivity also need further study.

\vspace{3ex}

\begin{acknowledgments}
We thank Yongkang Luo, Ningning Liu for insightful
discussions. This work was supported by the National Key R\&D
Program of the China (Grant Nos. 2016YFA0300402, 2019YFA0308602),
the National Science Foundation of China (Grant Nos. 11774305)
and the Fundamental Research Funds for the Central
Universities of China.
\end{acknowledgments}





%

\end{document}